%% file: main.tex
\documentclass[a4paper,11pt]{article}
\usepackage{pos}
\usepackage{slashed}
\usepackage{multirow}
\usepackage{subcaption}
\usepackage{hyperref}
\usepackage{subcaption} 
\usepackage{setspace}

\let\OLDthebibliography\thebibliography
\renewcommand\thebibliography[1]{
  \OLDthebibliography{#1}
  \setlength{\parskip}{0pt}
  \setlength{\itemsep}{0pt plus 0.3ex}
}

\input{definitions.tex}
\graphicspath{{./Figures/}}

\title{
Isospin breaking corrections to the hadronic vacuum \\ polarization with stochastic coordinate sampling}

\author[a,b]{Mattia Bruno}
\author[d]{Vera Gülpers}
\author[d]{Nils Hermansson-Truedsson}
\author[c]{Christoph Lehner}
\author*[c]{Julian Parrino}
\author[e]{J. Tobias Tsang}

\affiliation[a]{Dipartimento di Fisica ``Giuseppe Occhialini'', Universit\`a degli Studi di Milano-Bicocca,\\ Piazza della Scienza 3, 20126 Milan, Italy}
\affiliation[b]{Istituto Nazionale di Fisica Nucleare (INFN), Sezione di Milano-Bicocca,\\ Piazza della Scienza 3, 20126 Milan, Italy}
\affiliation[c]{Fakult\"at f\"ur Physik, Universit\"at Regensburg, Universit\"atsstra{\ss}e 31, 93040 Regensburg, Germany}
\affiliation[d]{ Higgs Centre for Theoretical Physics, School of Physics and Astronomy,\\ The University of Edinburgh, James Clerk Maxwell Building,
\\Peter Guthrie Tait Road,
Edinburgh,
EH9 3FD}
\affiliation[e]{Department of Mathematical Sciences, University of
Liverpool, \\Liverpool L69 3BX, United Kingdom}

\emailAdd{julian.parrino@ur.de}

\abstract{
In the recent Muon g-2 Theory Initiative white paper update, the hadronic vacuum polarization (HVP) contribution -- which dominates the theoretical uncertainty -- is evaluated as an average of different lattice QCD calculations. Since lattice simulations are mostly carried out in isospin symmetric QCD, corrections due to the mass difference of the up and down quarks and the coupling to photons have to be accounted for. These isospin breaking effects are of order 1\% and can be treated as corrections to the result for the HVP contribution in isospin symmetric QCD. In the current estimate of the HVP contribution, these effects are a large source of uncertainty due to the extensive computational cost to compute all occurring Wick contractions and degrading signal-to-noise behaviour especially for quark disconnected diagrams.

We present the current status of the calculation of isospin breaking corrections in the HVP contribution for the RBC/UKQCD collaborations. We use a dataset of
propagators computed using stochastic coordinate sampling (SCS)
to construct all necessary Wick contractions for the electromagnetic and strong isospin breaking effects. We employ different versions of QED on the lattice, such as QED$_L$, QED$_r$ and QED$_\infty$ to improve our estimate of finite-volume uncertainties.
}

\FullConference{%
The 42nd International Symposium on Lattice Field Theory (LATTICE2025),\\
2-8 November 2025,\\
Tata Institute of Fundamental Research, Mumbai, India
}

\begin{document}\let\oldref\ref

\addtocounter{page}{-1}
\maketitle

\input{intro}
\input{formalism}
\input{implementation}

\input{results}
\input{conclusion}

\input{ackno}

\bibliographystyle{apsrev4-1}
{\footnotesize

\bibliography{refs}

}

\end{document}

%% file: definitions.tex
\renewcommand{\vec}[1]{\boldsymbol{#1}}
\newcommand{\be}{\begin{equation}}
\newcommand{\ee}{\end{equation}}
\newcommand{\ba}{\begin{eqnarray}}
\newcommand{\ea}{\end{eqnarray}}

\newcommand{\tr}{{\rm Tr\,}}

\newcommand{\vev}[1]{\Big\langle #1 \Big\rangle_{U}}

%% file: intro.tex
\section{Introduction}
At the frontier of precision physics, the anomalous magnetic moment of the muon [$a_\mu =\frac12 (g-2)_\mu$] serves as one of the most stringent tests of the Standard Model (SM) of particle physics.
\cite{Jegerlehner:2017gek,Aoyama:2020ynm,Aliberti:2025beg}. With the new $a_\mu$ measurements at Fermilab~\cite{Aguillard:2025fij}, reaching a precision of  $127$ parts-per-billion (ppb)~\cite{Aguillard:2025fij}, the experimental precision now exceeds the prediction given by the Muon $g - 2$ Theory Initiative~\cite{Aliberti:2025beg} by a factor of four. It is therefore crucial to further develop the theoretical prediction, to match the experimental uncertainty.

In the SM, $a_\mu$ receives contributions from all fundamental interactions. While pure QED effects are by far the dominant contribution, several decades of improvements in the perturbative calculation of these effects have led to an astonishing precision of 1.7 ppb \cite{Aliberti:2025beg}. The dominant source of error is dictated by hadronic interactions, where the hadronic vacuum polarization (HVP) amounts to the largest part of the total uncertainty with an error of 523 ppb. This difficulty arises due to the non-perturbative nature of the strong interaction in the energy regime relevant for the determination of $a_\mu$. There are two complementary methods for the determination of the HVP contribution:  the dispersive approach, derived from unitarity and analyticity of the S-Matrix combined with experimental data from $e^+e^-$ to hadron decays, and lattice QCD, where quarks and gluon interactions can be simulated from first principles.
Within the dispersive approach there are significant discrepancies between different datasets that prohibit the determination of the HVP contribution with the desired precision.
There is currently a significant effort (see, e.g., \cite{Budassi:2026lmr}) to clarify the origin of the discrepancies.
On the other hand, lattice QCD has made significant progress in recent years, with several results determining the HVP contribution with approximately $1\%$ precision. Both facts led to the decision to adopt the lattice QCD average as the result for the HVP contribution in WP 2025 \cite{Aliberti:2025beg}.
It is also possible to replace hadronic $e^+e^-$ decay data with data obtained from hadronic $\tau$ decays if the needed isospin breaking (IB) corrections are reliably estimated \cite{Bruno:2018ono,Miranda:2020wdg,Colangelo:2025ivq}.

In lattice QCD, simulations are typically performed in isospin symmetric QCD, i.e., with equal masses of the up and down quark and without dynamical photons. Since these effects are of the order of $1\%$, it is necessary to understand and quantify these effects in the HVP contribution in order to match the experimental precision of $a_\mu$. While computationally challenging, lattice simulations with dynamically QCD+QED have been carried out successfully \cite{Borsanyi:2020mff,Boccaletti:2024guq}, it is also possible to understand isospin breaking corrections as an expansion -- the RM123 approach \cite{deDivitiis:2013xla} -- around isospin symmetric QCD (isoQCD), introducing correlation function that are higher order in the electromagnetic coupling $\alpha$, as well as the quark mass shifts $\Delta m_f$ \cite{blum:2018mom,Parrino:2025afq,Erb:2025nxk}. The diagrams that arise in this method are depicted in Figs.~\ref{fig:diags} and \ref{fig:sib_diagrams}.
In a first step, the isoQCD scheme needs to be selected for which we consider hadronic schemes \cite{Aliberti:2025beg,RBC:2024fic,RBC:2023pvn,blum:2018mom}.  In a second
step, the completion of the calculation of isospin breaking corrections to the HVP requires the determination of the mass splitting, e.g., using the mass difference of the neutral and charged kaon.

The inclusion of photons leads to additional infrared and ultraviolet divergences.  In the literature there are multiple regularization schemes, defining a lattice formulation of QED.  In this work, we use QED$_L$\cite{Hayakawa:2008an}, QED$_r$\cite{DiCarlo:2025uyj}, QED$_\infty$ \cite{Blum:2017cer,blum:2018mom,Chao:2020kwq,Parrino:2025afq} to
reliably estimate systematic effects from the finite simulation volume.
A complementary approach utilizes a Pauli-Villars regularized photon propagator in the continuum and infinite volume \cite{Biloshytskyi:2022ets,Erb:2025nxk}.

In these proceedings, we report on the current status of the calculation of IB corrections of the RBC/UKQCD collaborations, improving on the first calculation of the leading IB corrections in 2018 \cite{blum:2018mom}. 
We employ the RM123 method beyond electro quenched QCD+QED using stochastic coordinate sampling (SCS) of all-to-all propagators to calculate the necessary correlation functions. 
We show preliminary results for the blinded integrand for several contributions 

%% file: formalism.tex
\section{Methodology of isospin breaking corrections}
\label{sec:formalism}
The primary object that we are interested in, is the spatially summed correlator of two electromagnetic vector currents projected onto vanishing spatial momentum
\be 
\label{eq:spatsummedcor}
{\cal C}(t)\ = \frac{1}{3} \sum_{i=1,2,3} \int d^3\vec{z} \langle j_i(t,\vec{z})j_i(0) \rangle_{\rm QCD+QED}\,,
\ee 
where $j_\mu(x)=i\sum_{f}Q_f\bar{\psi}_f(x)\gamma_\mu \psi_f(x)$, for quark flavors $f=\{u,d,s,c,b,t\}$ with charges $Q_f=\{\frac{2}{3},-\frac{1}{3},-\frac{1}{3},\frac23,-\frac13,\frac23 \}$. 
The HVP contribution to $a_\mu$ in the time-momentum representation (TMR) \cite{Bernecker:2011gh}, is obtained from
\be 
\label{eq:a_mu_tmr}
a_\mu^{\rm HVP} = \Big(\frac{\alpha}{\pi}\Big)^2 \int_0^\infty dt \,f(t,m_\mu)\, {\cal C}(t),
\ee 
with an analytically known kernel function $f(t,m_\mu)$. In the RM123 approach \cite{deDivitiis:2013xla}, we expand the vector correlator in Eq. \eqref{eq:spatsummedcor}, around isoQCD
\ba 
{\cal C}(t)\ = {\cal C}^{\text{isoQCD}}(t)\ + {\cal C}^{\text{QED}}(t)\ + {\cal C}^{\text{SIB}}(t)\ + O\Big(\alpha^2,(\Delta m_f)^2, \alpha (\Delta m_f)\Big),
\ea 
where for the light-quark correlator in isoQCD, only the quark-connected (c) and disconnected diagram (d) contribute \cite{Aliberti:2025beg}
$
C^{\text{isoQCD}} (t) = \frac{5}{9}({\rm c}) - \frac{1}{9} ({\rm d})$.
The QED effects at $O(\alpha)$ are given  by
\ba 
C^{\text{QED}}(t) &:=&\alpha \left[  \frac{d}{d\alpha} \frac13 \sum_{i=1,2,3} \int d^3\vec{z} \langle j_i(t, \vec{z})j_i(0) \rangle_{\text{QCD+QED}}\right]_{\alpha=0,\Delta m_u=\ldots=\Delta m_t=0} \\
&=&-2\pi \alpha \frac13 \sum_{i=1,2,3} \int d^3\vec{z}  \int_{x,y}\delta_{\nu \rho}G(y,x) \langle j_i(t,\vec{z}) j_\nu(y) j_\rho(x) j_i(0) \rangle_{\text{isoQCD}},
\label{eq:C_QED}
\ea
where $\delta_{\nu \rho}G(y,x)$ is the photon propagator in Feynman gauge, and $\langle j_\mu(z) j_\nu(y) j_\rho(x) j_\sigma(0) \rangle_{\text{isoQCD}}$ is the four-point function evaluated in isoQCD. For greek indices occuring in pairs, we use the Einstein sum convention. The strong isospin breaking (SIB) effects are obtained from
\ba 
\label{eq:def_sib_contribution}
C^{\text{SIB}}(t):= \frac13 \sum_{f,i} \Delta m_f  \left[ \frac{d}{d(\Delta m_f)}\langle j_i(t,\vec{z})j_i(0) \rangle_{\rm QCD}\right]_{\Delta m_u=\ldots=\Delta m_t=0}
\ea

\begin{figure}
    \centering
    \includegraphics[width=0.9\linewidth]{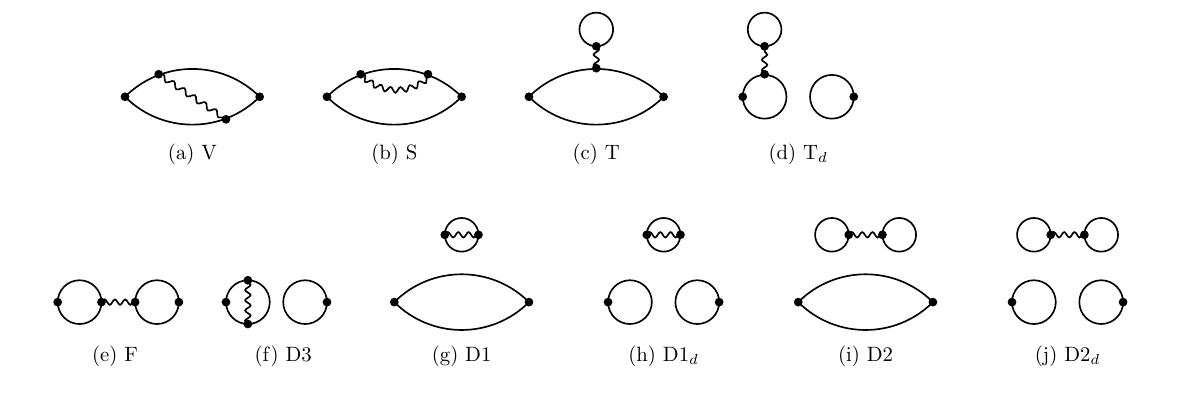}
    \caption{QED corrections to the HVP contribution, taken from Ref.~\cite{Gulpers:2018mim}. Gluons are not depicted, so that for example all quark-
disconnected diagrams are connected by gluons.}
    \label{fig:diags}
\end{figure}

Performing the Wick contractions in the light-quark sector, one obtains
\ba 
\label{eq:G1gamma_star}
C^{\text{QED}}(t)&=& 4\pi \alpha \Big\{ -\frac{17}{81}((\rm{V})+2(\rm{S}))+\frac{25}{162}(2(\rm{F})+(\rm{D1}))+\frac{14}{81}((\rm{T})+(\rm{D3})) \\
\nonumber && \hspace{2cm} -\frac{10}{81}(\rm{T}_d) -\frac{5}{162}((\rm{D1}_d)+(\rm{D2})) +\frac{1}{162}(\rm{D2}_d)\Big\}
\ea 
with diagrams depicted in Fig.~\ref{fig:diags}. In Eq.~\eqref{eq:G1gamma_star}, the diagram names (V), (S), (F), etc., are defined through traces of propagators and gamma structures summed over internal  vertices without the inclusion of charge factors or symmetry factors due to equivalent contractions.  This means, e.g.,
\ba 
\label{eq:def_V}
({\rm V})(t) := \frac13 \sum_{i=1,2,3} \int d^3\vec{z}  \int_{x,y}G(y,x)\langle\tr\{\gamma_i S(0;x)\gamma_\alpha S(x;\vec{z},t)\gamma_iS(\vec{z},t;y) \gamma_\alpha S(y;0)\} \rangle_{U}
\ea 
where $S(y;x)$ denotes the light-quark propagator from source position at $x$ to the sink position $y$ and $\langle \dots \rangle_U$ denote the average over gauge configurations in isoQCD. 
In Fig.~\ref{fig:diags} quark-disconnected diagrams such as (F) and (D1) are understood to be connected by gluons.  In order to avoid double counting with the next-to-leading-order HVP contribution \cite{blum:2018mom}, the vacuum expectation value (VEV) of individual quark-disconnected subdiagrams in diagram (F) needs to be subtracted, i.e.,
\ba 
\nonumber 
({\rm F} )(t) &=& \frac13 \sum_{i=1,2,3} \int d^3\vec{z} \int_{x,y}G(y,x)\Big[ \vev{\tr\Big\{\gamma_i S(0;x)\gamma_\alpha S(x;0)\Big\} \tr\Big\{ \gamma_\alpha S(y;t,\vec{z}) \gamma_i S(t,\vec{z};y)\Big\}} \\
&& \hspace{2.5cm} -\vev{\tr \Big\{\gamma_i S(0;x)\gamma_\alpha S(x;0)\Big\}}\vev{\tr\Big\{\gamma_\alpha S(y;t,\vec{z}) \gamma_i S(t,\vec{z};y)\Big\}}\Big]\,.
\label{eq:diag-F}
\ea 
A similar subtraction is also needed for the diagrams
(D1), (D1$_{\rm d}$), (D2), and (D2$_{\rm d}$) due to the normalization of the path integral.

\begin{figure}
    \centering
    \includegraphics[width=0.7\linewidth]{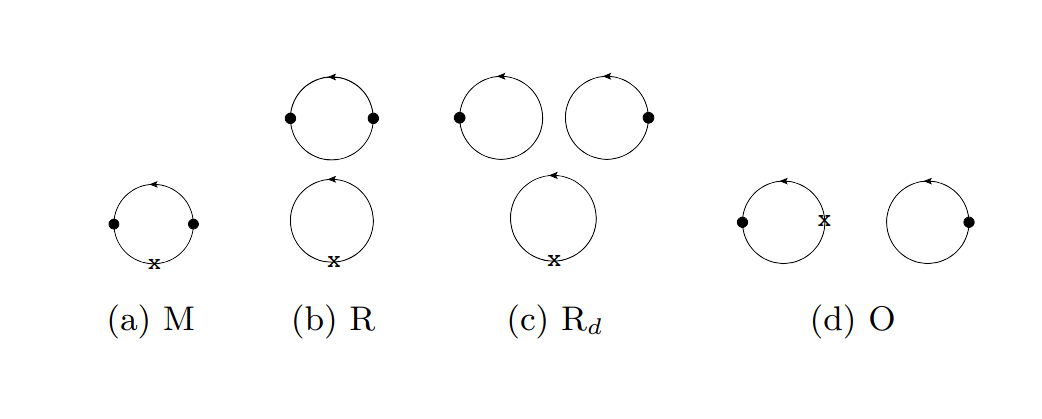}
    \caption{Strong isospin breaking diagrams, taken from Ref.~\cite{Gulpers:2018mim}}
    \label{fig:sib_diagrams}
\end{figure}

The strong isospin breaking corrections can be evaluated by writing the mass derivative with the insertion of a scalar operator.  For the light quark component the resulting diagrams, depicted in Fig.~\ref{fig:sib_diagrams} are given by  
\ba 
\label{eq:sib_contribution}
C^{\text{SIB}}(t)&:=& -\frac19(\Delta m_u+\Delta m_d) (R_d)+ (-\frac89 \Delta m_u-\frac29\Delta m_d)(M)\\
& & +(\frac49 \Delta m_u-\frac29\Delta m_d)(O)+\frac59(\Delta m_u+\Delta m_d) (R)\,,
\ea
where the mass shift parameters need to be determined from a separate calculation of hadronic quantities. A typical choice for a hadronic renormalization scheme in $N_f=2+1$ flavor QCD+QED \cite{blum:2018mom} proposes to fix the neutral pion and kaon mass $m_{\pi^0}$ and $m_{K^0}$, the mass difference of the neutral and charged kaon $\Delta (m_{K^0}-m_{K^\pm})$ and a mass of the $\Omega^{-}$ baryon $m_\Omega$, expanding naturally around the isospin symmetric scheme used in the RBC/UKQCD determination of the long-distance window quantity \cite{RBC:2024fic}. If a local vector current is used in the lattice determination of the IB corrections, it is furthermore required to compute the IB corrections to the renormalization factor $Z_V$ for a complete calculation.

%% file: implementation.tex
\section{Stochastic coordinate sampling}
\label{sec:implementation}
We perform calculations on three gauge ensembles, see table \ref{tab:ensembles}, generated by the RBC/UKQCD collaborations, which have been used in the determination of the HVP contribution at the isospin symmetric point \cite{RBC:2024fic}. For the computation of the four point functions, we use  stochastic coordinate sampling (SCS), analogous to the RBC/UKQCD  calculation of the hadronic light-by-light contribution \cite{Blum:2015gfa,Blum:2016lnc,Blum:2017cer,Blum:2019ugy,Blum:2023vlm}. We construct the contractions for diagram (V) from a dataset of  point-source propagators. For each ensemble these propagators have been precomputed for random sampled tuples of N source and M sink positions, given in table \ref{tab:ensembles}. Drawing propagators from this dataset, diagram (V) can then be calculated using the estimator
\ba 
\nonumber
&&\hspace{-0.7cm}({\rm{V}})(t):=\frac{ \zeta_{\text{V}} }{3} \langle \hspace{-0.3cm}\sum_{i,w_l,x_k,y_q, \vec{z}_j} \hspace{-0.3cm}G(y_q,x_k) \tr\{\gamma_i S(w_l;x_k)\gamma_\alpha S(x_k;\vec{z}_j,t_z)\gamma_i S(\vec{z}_j,t_z;y_q) \gamma_\alpha S(y_q;w_l)\} \Big|_{t=t_z-t_{w_l}} \hspace{-0.8cm}  \rangle_U . 
\\ 
\label{eq:def_V_implementation}
\ea 
The summation is performed over sampled coordinates $w_l,x_k,y_q, \vec{z}_j$, with $\zeta_{\text{V}} = \frac{(\frac{L^3T}{a^4})^3}{N_{w_l} N_{x_k} N_{y_q} N_{z_j}}$ being the correction factor to account for the number of samples used. 
The number of samples in specific coordinate-space subdomains can be adjusted, to study for example distance cuts between the $x$ and $y$ vertex, or to sample regions with larger contributions more densely.
For diagram (F), we use a similar dataset of stochastic source positions, without sparsening in the sink domain and computing directly the spatially summed two-point correlator for each source position $x_k$
\ba 
C_{\alpha \beta}(x_k;t) = \sum_{\vec{y}} \tr \{\gamma_\alpha S(x_k;y)\gamma_\beta S(y;x_k)\}\Big|_{t=t_y-t_{x_k}},
\ea 
where the full spatial volume sum is performed for $y$. For the VEV subtraction, i.e., the second line of Eq.~\eqref{eq:diag-F}, we subtract the gauge average of the two-point for each source position $x_k$
\ba 
\widetilde{C}_{\alpha \beta}(x_k;t)=C_{\alpha \beta}(x_k;t) - \langle \frac{1}{N_{w_l}}\sum_{w_l} C_{\alpha \beta}(w_l;t) \rangle_U.
\ea 
Diagram (F) is then obtained from 
\ba 
\label{eq:implementation_F}
({\rm{F}})(t)= \frac{ \zeta_{\text{F}} }{3}\sum_{i} \langle G(x_k,y_q) \sum_{x_k,y_q,t_w,t_z} \widetilde{C}_{\alpha i}(x_k;t_w) \widetilde{C}_{\alpha i}(y_q;t_z)\Big|_{t=t_z-t_w} \hspace{-0.6cm} \rangle_U ,
\ea 
with the corresponding correction factor for the number of samples taken into account $\zeta_{\text{F}}$. Analogous to Eq.~\eqref{eq:def_V_implementation}, it is instructive to study different subdomain in position-space, such as distance cuts between the $x$ and $y$ vertex. The all-mode-averaging procedure for each propagator can be applied in Eqs.~\eqref{eq:implementation_F} and \eqref{eq:def_V_implementation} by considering all different combinations of exact and sloppy solves as well as low modes and account for each contribution by modifying the correction factors $\zeta$.

For quark-disconnected diagrams that involve a tadpole such as diagrams (T), (D2), (D3),  we compute estimators of the tadpole field $
    T^{(f)}_x = \langle \psi_f(x)\bar\psi_f(x)\rangle
$
with quark flavor $f \in \{u,d,s\}$. Here, we employ a method similar to the computation for the quark-disconnected contribution in isoQCD \cite{Blum:2015you}.
We write the tadpole field in the following way 
$
T^{(ud)} - T^{(s)} = T^{(ud,\text{low})} + T^{(ud,\text{high})} - T^{(s)}
$
 with $T^{(ud,\text{high})} = T^{(ud)} - T^{(ud,\text{low})}$ and Dirac low-mode reconstruction $T^{(ud,\text{low})}$. We use multi-grid Lanczos \cite{Clark:2017wom} to compute the exact low-mode field $T^{(ud,\text{low})}_x$ on over 2000-5000 lowest preconditioned Dirac modes and estimate $T^{(ud,\text{high})} - T^{(s)}$ using a random sparse $Z_2$ grid.  The sparseness of the grid is crucial to suppress noise from unwanted contractions.  In practice a sparseness of $3^4$ is often sufficient for the ensembles of table \ref{tab:ensembles}, which means that the total field can be filled with $3^4=81$ propagator computations.

 For the photon propagator in lattice regularization, with $\hat{k}^2=\frac{4}{a^2}\sum_{\alpha=1}^4 \sin^2 \Big(\frac{a k_\alpha }{2}\Big)$, we utilize the commonly used QED$_L$ prescription \cite{Hayakawa:2008an}
$
G^{\text{QED}_L}(x,y) = \frac{1}{L^3 T}\sum_{k} e^{ik(x-y)} (1-\delta_{\vec k,0})(1/\hat{k}^2)$
as well as the recently introduced QED$_r$ variant \cite{DiCarlo:2025uyj}
$
 G^{\text{QED}_r}(x,y) =  \frac{1}{L^3 T} \sum_{k} e^{ik(x-y)} (1+\delta_{|\vec{k}|,2\pi/L}/6)(1-\delta_{\vec k,0})(1/\hat{k}^2)
$,
where the zero modes of the photon propagator are redistributed over the lowest non-zero modes. Here, the sum over $k$ is performed over all lattice momenta. We also study QED$_\infty$ by increasing the extent L of the QED box subsequently, e.g. QED$_{6L}$ in Fig.~\ref{fig:integrand_leading}, which in the limit $L\to \infty$ converges to the infinite volume photon propagator \cite{blum:2018mom} $
G^{\text{QED}_\infty}(x,y) = \int_{-\pi/a}^{\pi/a} \frac{d^4k}{(2\pi)^4} \frac{1}{\hat{k}^2} e^{i k x}\,
$.
\begin{table}
\caption{Parameters of the ensembles used in this work. All values are taken from Ref.~\cite{RBC:2024fic}.}
\label{tab:ensembles}
  \begin{tabular}{l|llllllll}
  ID & $a^{-1}$/GeV &  $L^3 \times T \times L_s/a^5$ & $m_\pi$/MeV & $m_K$/MeV & $m_\pi L$ & N (exact/sloppy) & M \\ \hline
  48I & $1.7312(28)$ &  $48^3 \times 96 \times 24$ & $139.32(30)$ & $499.44(88)$ & 3.9 & 64 / 2048 & \hspace{-0.8cm} 663552 \\
   C & $1.7312(28)$ &  $64^3 \times 128 \times 24$ &  $\approx 139$ & $\approx 499$ & 5.2 & 256 / 4096 & \hspace{-1cm}  1048576 \\\hline
  4   & $1.7312(28)$ &  $24^3 \times 48 \times 24$  &  $274.8(2.5)$ & $530.1(3.1)$ & 3.8  & 32 / 1024 & \hspace{-0.6cm} 41472\\
  \end{tabular}
\end{table}
As a systematic check for the implementation of the Wick contractions for all fourteen diagrams for the isospin breaking corrections to the HVP, we performed the following strategy: We computed the full all-to-all propagator on a single gauge configuration on a $4^3\times8$ lattice. Because of the small lattice extent all diagrams can be computed from the all-to-all propagator. In this way, we obtained a reference implementation, for all diagrams, for all possible combinations of external operators and the different photon implementations that we used. The lattice QCD code that implements the stochastic sampling technique in GPT \cite{GPT} is checked against this reference implementation ensuring convergence against the full all-to-all propagator with increasing number of stochastic propagators.

%% file: results.tex
\section{Preliminary results}
In this section, we discuss preliminary results for some of the contribution from different diagrams to Eq.~\eqref{eq:G1gamma_star}. Our goal is to demonstrate the viability of our sampling procedure. So, in order to not spoil a potential future blinding strategy, we only investigate the integrand of Eq. \eqref{eq:a_mu_tmr} using a simple $t^4$ weight function, which behaves similar as the correct TMR kernel in the full integration region and multiply a blinding factor.
When comparing the contribution from different diagrams, we however consider multiplicative chargefactors indicated in the corresponding figure, whenever this is relevant.

\subsection{Connected contributions}
\begin{figure}
        \begin{subfigure}{0.49\textwidth}
            \centering
            \includegraphics[width=0.99\textwidth]{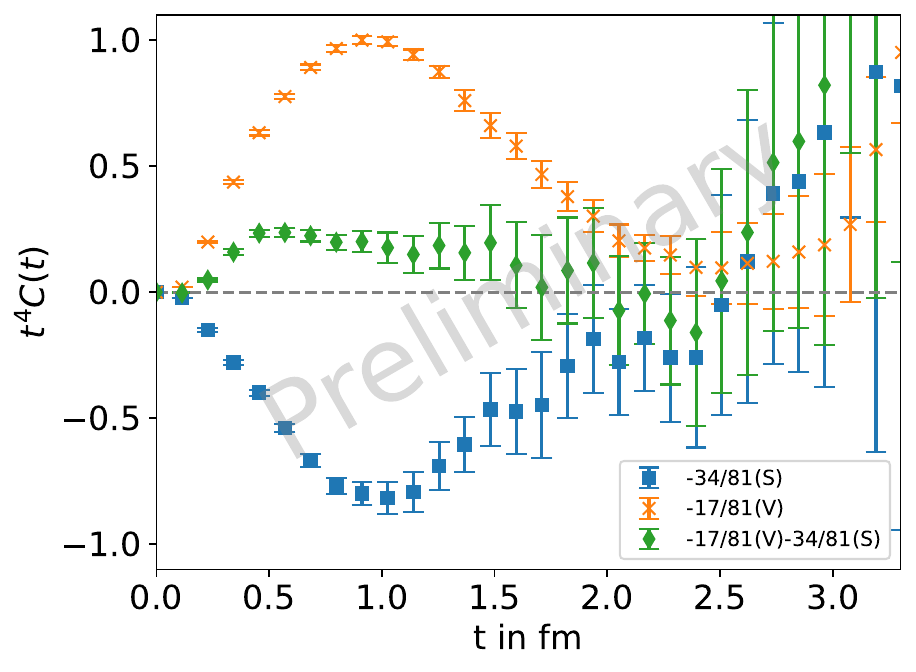}
            \subcaption{connected contributions}
            \label{fig:integrand_connected}
        \end{subfigure}
        \begin{subfigure}{0.49\textwidth}
            \centering
            \includegraphics[width=0.99\textwidth]{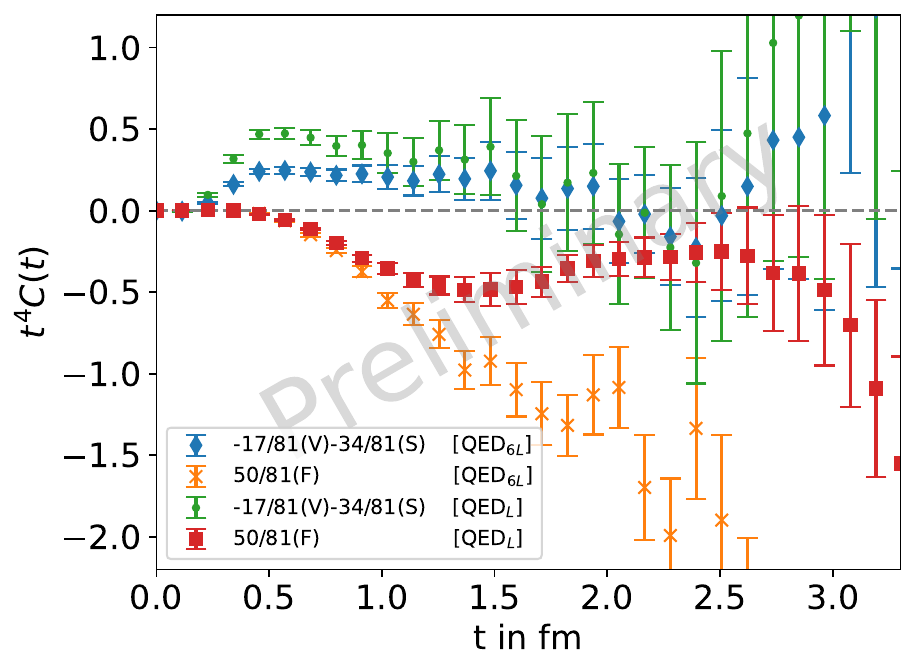}
            \subcaption{connected and leading disconnected contributions}
            \label{fig:integrand_leading}
        \end{subfigure}
        \begin{subfigure}{0.49\textwidth}
            \centering
            \includegraphics[width=0.99\textwidth]{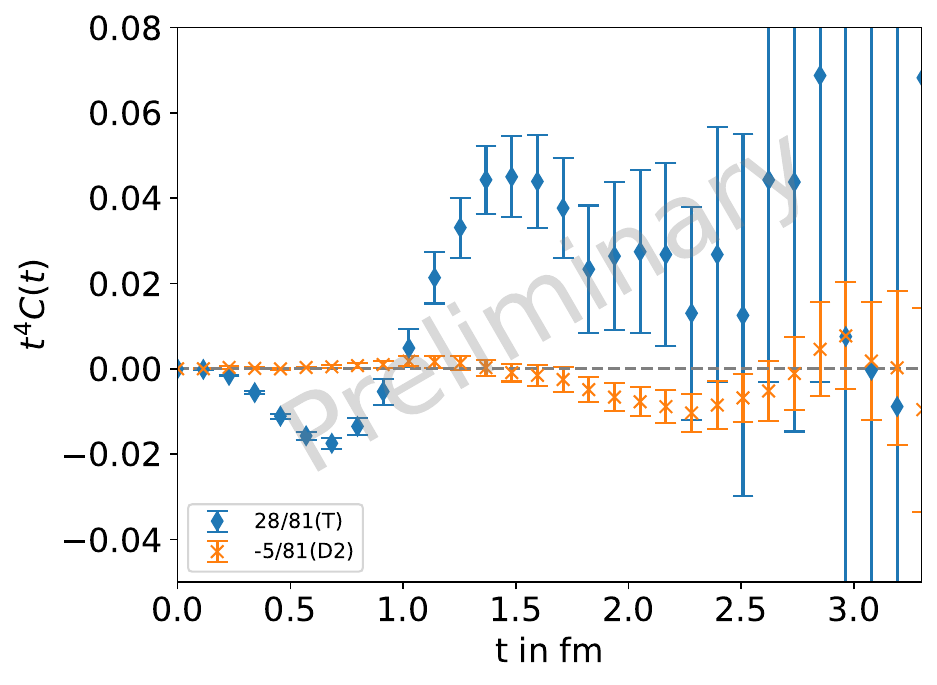}
            \subcaption{Tadpole contributions (T) and (D2)}
            \label{fig:integrand_tadpole}
        \end{subfigure}
        \begin{subfigure}{0.02\textwidth}
        \quad
        \end{subfigure}
        \begin{subfigure}{0.49\textwidth}
            \centering
            \hspace{1.4cm}
            \includegraphics[width=0.95\textwidth]{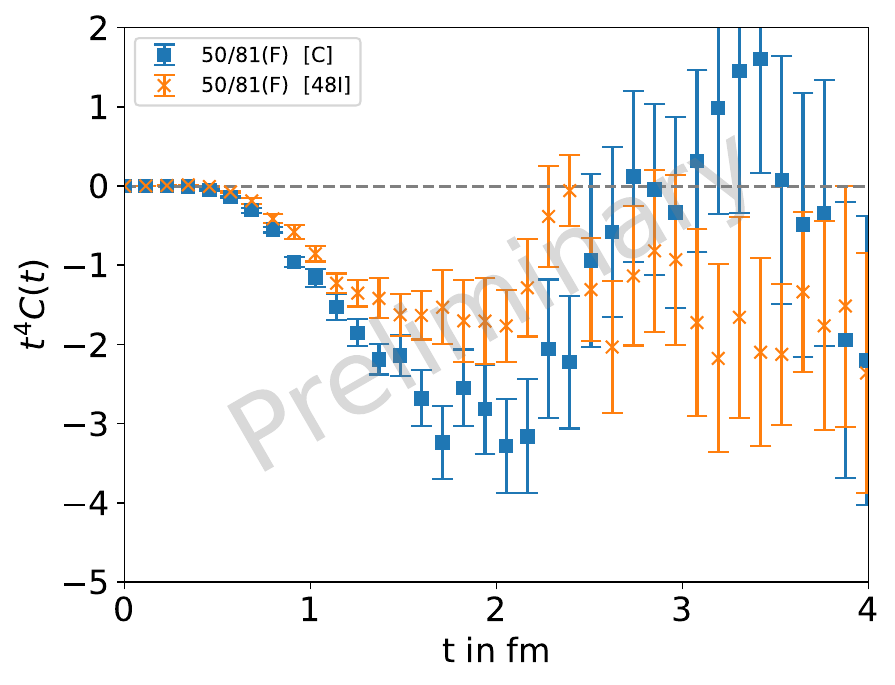}
            \subcaption{Disconnected contribution (F) at the physical point}
            \label{fig:integrand_diagF}
        \end{subfigure}
        \caption{Integrands for several diagrams, contributing to Eq.~\eqref{eq:G1gamma_star}. For each plot, we normalize the y-axis by the value of the peak of the contribution from the diagram (V) in QED$_L$ on the ensemble 4.}
        \label{fig:integrands}
\end{figure}

We display the integrands for the connected diagrams (V) and (S) on ensemble 4 in Fig.~\ref{fig:integrand_connected}. We note a strong cancellation in the sum $(V)+2(S)$, that appears in Eq. \eqref{eq:G1gamma_star}, as previously seen in the earlier work by the RBC/UKQCD collaboration \cite{blum:2018mom} and independently verified by the Mainz collaboration \cite{Erb:2025nxk}. The statistical uncertainty of the combined integrand is dominated by the self energy diagram (S), where the region above $1.5$fm contributes significantly. To mitigate this problem, we proposed a reconstruction of the tail of the correlator, by studying the intermediate $\pi\gamma$ and $\rho$ state, described in Ref. \cite{Lehner:2025qrl}. This method leads to a significant reduction of the uncertainty of the noisy diagram (S).

\subsection{Disconnected contributions}
As previously shown \cite{blum:2018mom, Djukanovic:2024cmq,  Parrino:2025afq} the diagram (F) contributes significantly and with an opposite sign as the connected contributions to Eq.~\eqref{eq:G1gamma_star}, leading to a large cancellation in the total contribution, see Fig.~\ref{fig:integrand_leading}. For QED$_L$, the subtraction of the VEV of the individual quark-disconnected subdiagrams, i.e. the second line of Eq.~\eqref{eq:diag-F} is not strictly necessary, since it is exactly zero in that case. 
However, we observe that performing the subtraction on each summand of the SCS data in Eq.~\eqref{eq:implementation_F} leads to a significant reduction of the statistical error. Investigating cuts between the distance of the internal vertices $|x-y|$ of diagram (F), we also observed that in the case of QED$_L$ the majority of the contribution is contained within $0.7$fm, which is, however, not the case for QED$_\infty$. In Fig.~\ref{fig:integrands}, we plot the combined contribution from all possible separations $|x-y|$.
We also study the volume dependence at the physical pion mass comparing the two ensembles 48I and C in Fig.~\ref{fig:integrand_diagF}. Similarly to the diagram (S), the problem of large
noise-to-signal ratio in the tail of the contribution is mitigated by reconstructing the tail with asymptotic intermediate states; see Ref.~\cite{Lehner:2025qrl}.

We compute all disconnected contributions involving multiple tadpoles fields using the methods described in sec.~\ref{sec:implementation}. A plot of the integrand for the diagrams (T) and (D2) is shown in Fig.~\ref{fig:integrand_tadpole}, demonstrating that the integrands for both contributions can be resolved with sufficient accuracy. However, both diagrams are sub-leading compared to the connected contributions.

\subsection{Strong isospin breaking contributions}
We depict the integrand for the SIB diagrams on ensemble 4 in Fig.~\ref{fig:integrand_sib}. Although a straightforward comparison to the QED diagrams can only be obtained after determining the quark mass shifts $\Delta m_u$, $\Delta m_d$, we note that these diagrams have a similar signal-to-noise problem in the large $t$ regime. We currently therefore perform additional computations using multiple different valence quark masses to obtain the mass derivative in Eq.~\eqref{eq:def_sib_contribution} through finite differences.

\subsection{Kaon mass splitting}
\begin{figure}
        \begin{subfigure}{0.49\textwidth}
            \centering
            \includegraphics[width=0.99\textwidth]{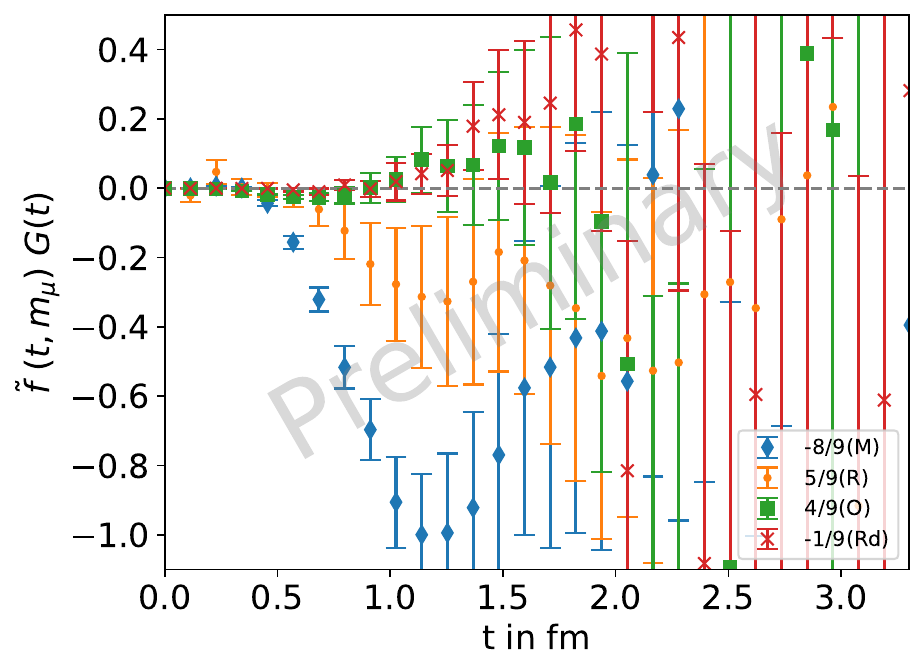}
            \subcaption{}
            \label{fig:integrand_sib}
        \end{subfigure}
        \begin{subfigure}{0.49\textwidth}
            \centering
            \includegraphics[width=0.99\textwidth]{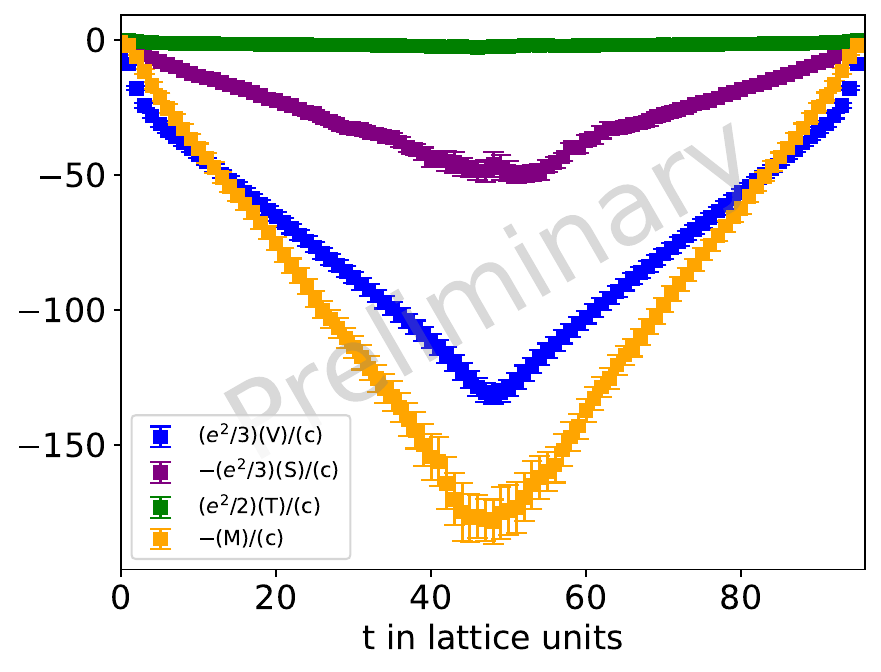}
            \subcaption{}
            \label{fig:kaon_mass_splitting}
        \end{subfigure}
        \caption{(a): Integrand for the SIB diagrams. The y-axis is normalized to the absolute height of the peak of diagram M. (b): Correlator ratios for the diagrams (V), (S), (T), and (M) contributing to $\Delta(m_{K^0}-m_{K^\pm})$ on the ensemble 48I.}
        \label{fig:2+2a_results}
\end{figure}

As outlined in section \ref{sec:formalism}, it is necessary to include an additional hadronic quantity associated with isospin breaking in addition to the hadronic scales used to define isoQCD. A first target in that regard is the mass difference between the neutral and charged kaon $\Delta(m_{K^0}-m_{K^\pm})$ that can be obtained from a subset of the same diagrams, as shown in Figs.~\ref{fig:diags} and \ref{fig:sib_diagrams} 
\ba 
\label{eq:kaon_mass_splitting}
e^2( Q_d - Q_u)Q_s \frac{({\rm V})}{({\rm c})}
+e^2(Q_d^2 - Q_u^2)\frac{({\rm S})}{({\rm c})}
+ e^2(Q_u-Q_d)\sum_f {Q_f} \frac{({\rm T})}{({\rm c})} 
-(m_d-m_u) \frac{({\rm M})}{({\rm c})},
\ea 
where $(c)$ is the connected contribution without photons.
Here, we consider pseudoscalar insertions for the external operators with quark-line directly from source to sink being the strange quark propagator.
For an infinitely large time extent $\Delta(m_{K^0}-m_{K^\pm})$ can be directly obtained by the time derivative of Eq.~\eqref{eq:kaon_mass_splitting} in the limit $t\rightarrow \infty$, however on a finite box Eq.~\eqref{eq:kaon_mass_splitting} can be fit to the expected functional form \cite{deDivitiis:2013xla}. The ratios of the contributing diagrams and the leading order are depicted in Fig.~\ref{fig:kaon_mass_splitting}. Here, we observe again that diagram (T) is subleading compared to the other contributions.

%% file: conclusion.tex
\section{Conclusion}
We have demonstrated the viability of our stochastic coordinate sampling approach to the computation of the IB effects for the HVP contribution, as well as the kaon mass difference. Supplemented by our long-distance reconstruction method \cite{Lehner:2025qrl} this approach is promising to reach the precision goal for required for a competitive determination of the HVP contribution. 
We are currently increasing the dataset to include more ensembles at the physical point with smaller lattice spacing, i.e., 64I and 96I that were also used in the determination of the long-distance window contribution in isoQCD \cite{RBC:2024fic}.
Additionally, we perform valence quark studies, as well as dynamical QED simulations to further investigate the contribution from noisy quark-disconnected contributions such as (D1) and (D1d).

%% file: ackno.tex
\vspace{0.3cm}
\tiny

\noindent{\textbf{Acknowledgements:\\}}
We thank our colleagues of the RBC and UKQCD collaborations for many valuable discussions and joint efforts over the years.
The authors gratefully acknowledge the Gauss Centre for Supercomputing
e.V. (www.gauss-centre.eu) for funding this project by providing
computing time on the GCS Supercomputer JUWELS at Jülich
Supercomputing Centre (JSC).  We acknowledge the EuroHPC Joint
Undertaking for awarding this project access to the EuroHPC
supercomputer LUMI, hosted by CSC (Finland) and the LUMI consortium
through a EuroHPC Extreme Scale Access call as well as the EuroHPC
supercomputer LEONARDO, hosted by CINECA (Italy).  An award of
computer time was provided by the ASCR Leadership Computing Challenge
(ALCC) and Innovative and Novel Computational Impact on Theory and
Experiment (INCITE) programs. This research used resources of the
Argonne Leadership Computing Facility, which is a DOE Office of
Science User Facility supported under contract DE-AC02-06CH11357. This
research also used resources of the Oak Ridge Leadership Computing
Facility, which is a DOE Office of Science User Facility supported
under Contract DE-AC05-00OR22725.  This research used resources of the
National Energy Research Scientific Computing Center (NERSC), a
U.S. Department of Energy Office of Science User Facility located at
Lawrence Berkeley National Laboratory, operated under Contract
No.~DE-AC02-05CH11231 using NERSC award NESAP m1759 for 2020. This
work used the DiRAC Blue Gene Q Shared Petaflop system at the
University of Edinburgh, operated by the Edinburgh Parallel Computing
Centre on behalf of the STFC DiRAC HPC Facility
(www.dirac.ac.uk). This equipment was funded by BIS National
E-infrastructure capital grant ST/K000411/1, STFC capital grant
ST/H008845/1, and STFC DiRAC Operations grants ST/K005804/1 and
ST/K005790/1. DiRAC is part of the National E-Infrastructure.  We
gratefully acknowledge disk and tape storage provided by USQCD and by
the University of Regensburg with support from the DFG.
The lattice data analyzed in this project was generated using GPT
\cite{GPT} and Grid \cite{GRID,Boyle:2016lbp,Yamaguchi:2022feu}.
This work is (partially) supported by ICSC – Centro Nazionale di Ricerca
in High Performance Computing, Big Data and Quantum Computing, funded by
European Union – NextGenerationEU. N.~H.-T.~is supported by the UK Research and Innovation, Science and Technology Facilities Council, grant number UKRI2426.